\documentclass{PoS}
\usepackage{epsfig}

\newcommand{\beq}{\begin{align*}}
\newcommand{\eeq}{\end{align*}}

\title{
\begin{picture}(0,0)(0,0)%
   \put(350,75){\makebox(0,0)[l]{\textnormal{\normalsize CHIBA-EP-179}}}%
   \put(350,65){\makebox(0,0)[l]{\textnormal{\normalsize KEK Preprint 2009-27}}}%
\end{picture}%
Gauge-independent derivation of "Abelian" dominance and magnetic monopole dominance 
in the string tension}

\ShortTitle{Gauge-independent derivation of "abelian" dominance and magnetic 
monopole dominance in the string tension}

\author{
   S.~Kato$^a$\thanks{Speaker at the conference(kato@takamatsu-nct.ac.jp).},  
   K.-I.~Kondo$^b$, 
   A.~Shibata$^c$, 
   T.~Shinohara$^b$   
   and S.Ito$^d$\\
\llap{$^a$}Takamatsu National College of Technology, Takamatsu City 761-8058, Japan\\
\llap{$^b$}Department of Physics, Graduate School of Science, Chiba University, Chiba 263-8522, Japan\\
\llap{$^c$}Computing Research Center, High Energy Accelerator Research Organization (KEK), Tsukuba 305-0801, Japan\\
\llap{$^d$}Nagano National College of Technology, 716 Tokuma, Nagano 381-8550, Japan
}

\abstract{
Recently, we have developed a reformulation of the lattice Yang-Mills theory based on
the change of variables a la Cho-Faddeev-Niemi combined with a non-Abelian Stokes theorem.
In this talk, we give a new procedure (called reduction) for obtaining the
color field which plays the central role in this reformulation.
In the 4D SU(2) lattice Yang-Mills theory, we confirm the
gauge-independent "abelian" dominance and gauge-independent
magnetic-monopole dominance in the string tension extracted from the
Wilson loop in the fundamental representation. 
}

\FullConference{The XXVII International Symposium on Lattice Field Theory\\
July 26-31, 2009\\
Peking University, Beijing, China}

\begin{document}

\section{Introduction}

It is interesting to study color confinement mechanism in Quantum Chromodynamics (QCD).
The dual superconductor scenario of the QCD vacuum may be a promising candidate for that
mechanism. In particular, it is known that the string tension calculated from the Abelian
and monopole parts reproduce well the original one, once we perform an Abelian projection
in Maximally Abelian (MA) gauge. It is so-called "Abelian and monopole dominance". But 
it has been diffcult to see these phenomena in any other gauges. 

Recently, we have demonstrated that the gauge-invariant magnetic monopole  can be constructed 
in the pure Yang-Mills theory without any fundamental scalar field. 
The success is achieved based on a new viewpoint proposed in \cite{KMS05} for the 
non-linear change of variables (NLCV), which was called Cho--Faddeev--Niemi (CFN) decomposition 
\cite{Cho80}\cite{FN98}, see also \cite{Shabanov99}. 
We have found that the magnetic charge of our lattice magnetic monopole is perfectly quantized.
Moreover, we have confirmed dominance of our magnetic monopole in the string tension\cite{IKKMSS06}, 
while it was first shown in \cite{SNW94} in the conventional MA gauge \cite{KLSW87}. 
Therefore we can show the gauge invariance of the dual superconductor scenario of the QCD vacuum.

In this talk, we summarize the recent results on a lattice formulation of Yang-Mills theory 
based on NLCV and a gauge-independent derivation of "Abelian" dominance and magnetic monopole dominance 
in the string tension. We restrict the following argument to SU(2) gauge group, for simplicity,
although the formulation has been extended to SU(N) gauge group \cite{KSM08}.

\section{Non-Linear change of variables (NLCV) in lattice SU(2) Yang-Mills theory}

We have proposed a natural and useful lattice formulation of  the NLCV
 in Yang-Mills theory corresponding to the CFN decomposition \cite{Cho80,FN98}. 
It is a minimum requirement that such a lattice formulation must reproduce  the continuum counterparts 
in the naive continuum limit. 
 
On a lattice, we introduce the site variable ${\bf n}_{x}$ , 
in addition to the original link variable $U_{x,\mu}$ which is related to the gauge 
potential ${\bf A}_\mu(x)$ in a naive way: 
\footnote{
In general, the argument of the exponential in (\ref{def-U}) is the line integral of a gauge potential 
along a link from $x$ to $x+\mu$.
 Note also that we define a color vector field 
${\bf n}(x) :=n^A(x){\sigma}_A/2$ in the continuum, while ${\bf n}_x  :=n_x^A\sigma_A$ on the lattice for convenience,
where $\sigma_A$ $(A=1,2,3)$ are Pauli matrices.
}
\begin{eqnarray}
U_{x,\mu} = \exp( -i \epsilon g {\bf A}_\mu(x)) , 
\label{def-U}
\end{eqnarray}
where $\epsilon$ is the lattice spacing and  $g$ is the coupling constant. Here  ${\bf n}_{x}$ is 
Hermitian, ${\bf n}_{x}^\dagger={\bf n}_{x}$, and $U_{x,\mu}$ is unitary, $U_{x,\mu}^\dagger=U_{x,\mu}^{-1}$.   
We call ${\bf n}_x$ a color unit vector field, since it is used to specify only the color direction in the color space at 
each space-time point and its magnitude is irrelevant (${\bf n}_{x}^2=1$).

The link variable $U_{x,\mu}$ and the site variable ${\bf n}_{x}$ transform under the gauge 
transformation II \cite{KMS05} as
\begin{eqnarray}
  U_{x,\mu} \rightarrow \Omega_{x} U_{x,\mu} \Omega_{x+\mu}^\dagger = U_{x,\mu}' , \quad
  {\bf n}_{x} \rightarrow \Omega_{x} {\bf n}_{x} \Omega_{x}^\dagger = {\bf n}_{x}' .
\end{eqnarray}

Then the link variable $U_{x,\mu}$ is decomposed into two parts:
\begin{eqnarray}
U_{x,\mu} = X_{x,\mu}V_{x,\mu} \in SU(2), \quad X_{x,\mu},V_{x,\mu}\in SU(2)
\label{NLCV-1}
\end{eqnarray}
in such a way that the color field ${\bf n}_{x}$ is covariantly 
constant in the background field $V_{x,\mu}$:
\begin{eqnarray}
 {\bf n}_{x} V_{x,\mu}  = V_{x,\mu} {\bf n}_{x+\mu} ,
 \label{Lcc}
\end{eqnarray}
and that the remaining field $X_{x,\mu}$ is perpendicular to all color fields ${\bf n}_{x}$ :
\begin{equation}
  {\rm tr}({\bf n}_{x} U_{x,\mu} V_{x,\mu}^\dagger) 
  = 0 .
  \label{cond2m}
\end{equation}
 Both conditions must be imposed to determine $V_{x,\mu}$ for a given set of ${\bf n}_{x}$ and 
$U_{x,\mu}$. 
By solving the {\bf defining equation} (\ref{Lcc}) and (\ref{cond2m}),
the link variable $V_{x,\mu}$ is obtained up to an overall normalization constant 
in terms of the site variable ${\bf n}_{x}$ and the original link variable 
$U_{x,\mu}$:
\begin{eqnarray}
  \tilde{V}_{x,\mu} = \tilde{V}_{x,\mu}[U,{\bf n}] 
  = U_{x,\mu} +  {\bf n}_{x} U_{x,\mu} {\bf n}_{x+\mu} .
  \label{sol}
\end{eqnarray}

Finally, the special unitary link variable $V_{x,\mu}[U,{\bf n}]$ is obtained after the normalization: 
\begin{eqnarray} 
V_{x,\mu} = 
V_{x,\mu}[U,{\bf n}] := 
 \tilde{V}_{x,\mu}/\sqrt{\frac{1}{2}{\rm tr} [\tilde{V}_{x,\mu}^{\dagger}\tilde{V}_{x,\mu}]} .
\label{cfn-mono-4}
\end{eqnarray}
It is easy to show that the naive continuum limit $\epsilon \rightarrow 0$ 
of the link variable $V_{x,\mu} = \exp (-i\epsilon g {\bf V}_\mu(x))$ reduces to 
\begin{eqnarray} 
{\bf V}_{\mu}(x) = (n^A(x)A_{\mu}^A(x)){\bf n}(x)
+\frac{1}{g}\partial_{\mu}{\bf n}(x)\times {\bf n}(x),
\label{cfn-conti-7} 
\end{eqnarray}
which is nothing but the continuum expression of CFN variable. 
Note that the $V_{x,\mu}$ transforms like a usual link variable under the gauge 
transformation II as
\begin{eqnarray}
  V_{x,\mu} \rightarrow \Omega_{x} V_{x,\mu} \Omega_{x+\mu}^\dagger = V_{x,\mu}' .
\end{eqnarray}

Therefore, we can define the {\it gauge-invariant flux}, $\bar{\Theta}_P[U,{\bf n}]$,
(plaquette variable) by
\begin{eqnarray} 
\bar{\Theta}_{x,\mu\nu}[U,{\bf n}] := \epsilon^{-2}
{\rm arg} ( {\rm tr} \{({\bf 1}+{\bf n}_x)V_{x,\mu}V_{x+\hat{\mu},\nu}
V_{x+\nu,\mu}^{\dagger}V_{x,\nu}^{\dagger} \}/{\rm tr}({\bf 1})) .
\label{cfn-mono-5}
\end{eqnarray}
It is also shown that the naive continuum limit of (\ref{cfn-mono-5}) reduces to
the gauge-invariant field strength;
\begin{eqnarray} 
\bar{\Theta}_{x,\mu\nu} \simeq
\partial_{\mu}(n^A(x)A_{\nu}^A(x))-\partial_{\nu}(n^A(x)A_{\mu}^A(x))
+ g^{-1} {\bf n}\cdot (\partial_{\mu}{\bf n}\times\partial_{\nu}{\bf n})
=\frac{-1}{2}{\rm tr}(2{\bf n}F_{\mu\nu}[V]),
\label{cfn-fs} 
\end{eqnarray}
which plays the similar role that 'tHooft--Polyakov tensor played in describing the 
magnetic monopole in Georgi--Glashow model. 

It has been shown that the SU(2) master Yang-Mills theory written in terms of 
$U_{x,\mu}$ and ${\bf n}_x$ has the enlarged local gauge 
symmetry $\tilde{G}^{\omega,\theta}_{local}=SU(2)_{local}^{\omega} \times [SU(2)/U(1)]_{local}^{\theta}$ 
larger than the local gauge symmetry $SU(2)_{local}^{\omega}$ in the original 
Yang--Mills theory\cite{KMS05}. 
  In order to eliminate the extra degrees of freedom in the enlarged local gauge symmetry 
$\tilde{G}^{\omega,\theta}_{local}$, 
we must impose sufficient number of constraints, which we call the {\bf reduction condition}. 
We find that the reduction condition is given by minimizing the following functional
\begin{eqnarray}
F_R[{\bf n},U] &=& \sum_{x,\mu} \left( 1- 
                   {\rm tr}({\bf n}_xU_{x,\mu}{\bf n}_{x+\hat{\mu}}U_{x,\mu}^{\dagger})
                   / {\rm tr}({\bf 1}) \right)
\end{eqnarray}
with respect to the color vector fields $\{{\bf n}_x\}$ for given link variables $\{U_{x,\mu}\}$.
Thus color vector field ${\bf n}_x$ is determined by ${\bf n}_x={\bf n}_x^*$ in such a way that
\begin{eqnarray}
{\bf min}_{{\bf n}} F_R[{\bf n},U] = F_R[{\bf n}^*, U].
\label{reduction_n}
\end{eqnarray}

The functional $F_R$ can be rewritten in the following way:
\begin{eqnarray}
F_R[{\bf n},U] 
= \sum_{<x,y>} (1-J^{AB}_{x,y}[U]n^A_xn^B_y), \quad 
J^{AB}_{x,y}[U] = {\rm tr}(\sigma^A U_{x,\mu}\sigma^B U_{x,\mu}^{\dagger})/{\rm tr}({\bf 1})
\end{eqnarray}
Therefore, the functional $F_R$ can be regarded as the spin-glass system.

We solve the stationaly condition
\begin{eqnarray}
\partial F_R[{\bf n},U]/\partial n^A_x =0
\end{eqnarray}
in order to minimize the functional $F_R$.
Note that there exist local minima which satisfy this condition. 
The overrelaxation method should be used in order to approach the global minimum more rapidly.

\section{Gauge-independent "Abelian" dominance in the Wilson loop}

Two of us\cite{KS08} discussed a gauge-independent definition of Abelian dominance in the Wilson loop operator and a 
constructive derivation of the Abelian dominance through a non-Abelian Stokes theorem via 
lattice regularization.

First, we insert the complete set of coherent states $|\xi_x,\Lambda>$ to the Wilson loop operator
$W_f[C]$ at every site $x$ on the loop $C$ to obtain
\begin{eqnarray}
W_f[C] = {\rm tr}({\cal P}\prod_{l\in C}U_l)/{\rm tr}({\bf 1}) 
       = \prod_{x\in C}\int d\mu(\xi_x)\prod_{l=<x,x+a\hat{\mu}>\in C}
           < \xi_x,\Lambda|U_l|\xi_{x+a\hat{\mu}},\Lambda  > .
\end{eqnarray}

Second, we consider decomposing the link field $U_{x,\mu}$ given by eq.(\ref{NLCV-1}), and we impose 
two requirements: 
\begin{itemize}
\item[(I)]  $\xi_x^{\dagger}V_{x,\mu}\xi_{x+a\hat{\mu}}\in \tilde{{\cal H}}=U(1)$ 
\item[(II)] $\rho_C[X,\xi] \equiv \prod_{l=<x,x+a\hat{\mu}>\in C}<\xi_x,\Lambda|X_{x,\mu}|\xi_x,\Lambda>=const.$
\end{itemize}
Here, $\tilde{{\cal H}}$ is a stability group of the gauge group $G=SU(2)$, and $\xi \in G/\tilde{\cal H}$.
Under the decomposition (\ref{NLCV-1}), the full Wilson loop reads
\begin{eqnarray}
W_f[C] = \prod_{x\in C}\int d\mu(\xi_x) \rho_C[X,\xi]
         \prod_{l=<x,x+a\hat{\mu}>\in C}
           <\xi_x,\Lambda|V_{x,\mu}|\xi_{x+a\hat{\mu}},\Lambda>.
\end{eqnarray}

It is shown that only a diagonal element of $<\xi_x,\Lambda|V_{x,\mu}|\xi_{x+a\hat{\mu}},\Lambda>$ 
contributes to $W_f[C]$ if and only if the requirement (I) is satisfied.  
Physically, requirment (I) is a condition for the field strength $F_{\mu\nu}[V](x)$ of the 
restricted field ${\bf V}_{\mu}(x)$ to have only the {\it Abelian part} proportional to the color field ${\bf n}(x)$
at the spacetime point x.

The requirement (II) allows us to factor out $\rho_C[X,\xi]$ and we obtain
\begin{eqnarray}
W_f[C] \simeq const. W_{abel}[C] = const.(\prod_{x\in C}\int d\mu(\xi_x)) 
         \prod_{l=<x,x+a\hat{\mu}>\in C} <\xi_x,\Lambda|V_{x,\mu}|\xi_{x+a\hat{\mu}},\Lambda> 
\label{NAST-1}
\end{eqnarray}

{\it It can be shown\cite{KS08} that requirement (I) and (II) are equivalent to defining equation (\ref{Lcc}) and (\ref{cond2m}) 
which are able to uniquely determine the NLCV.}
Therefore, (\ref{NAST-1}) suggest that $W_f[C]$ agrees with the restricted one $W_{Abel}[C]$ up to 
a constant factor in NLCV; 
\begin{eqnarray}
\left< W_f[C]\right> \simeq \left< W_{Abel}[C]\right> 
= \left< {\rm tr}(\prod_{l\in C}V_l)/{\rm tr}({\bf 1})\right> .
\end{eqnarray}
Since the restricted field ${\bf V}_{\mu}(x)$ is defined in a gauge-covariant and gauge independent way, 
we have obtained a gauge-invariant (and gauge-independent) definition of the "Abelian" 
dominance.

\section{Gauge-independent Monopole dominance in the Wilson loop}

We construct the gauge-invariant field strength (\ref{cfn-mono-5}) to extract  configurations of 
the (integer-valued) magnetic monopole current $\{k_{x,\mu}\}$  defined by
\begin{eqnarray}
k_{x,\mu}=
-\frac{1}{4\pi}{\varepsilon}_{\mu\nu\rho\sigma}
\partial_{\nu}\bar{\Theta}_{x+\mu,\rho\sigma} .
\label{cfn-conti-20}
\end{eqnarray}
This definition agrees with our definition of the monopole in the continuum limit (divided by $2\pi$).

In order to study the monopole dominance in the string tension, we proceed to estimate the 
magnetic monopole contribution $\left< W_m(C) \right>$ to the Wilson loop average, i.e., 
the expectation value of the Wilson loop operator $\left< W_f(C) \right>$.  
We define the magnetic part $W_m(C)$ of the Wilson loop operator $W_f(C)$ as the contribution 
from the monopole current $k_{x,\mu}$ to the Wilson loop operator:
\footnote{
The Wilson loop operator $W_f(C)$ is decomposed into the magnetic part $W_m(C)$ and the electric 
part $W_e(C)$, which is derived from the non-Abelian Stokes theorem, see \cite{Kondo00}.
In this talk, we do not calculate the electric contribution $\left< W_e(C) \right>$ where 
$W_e(C)$ is expressed by the electric current $j_{\mu}=\partial_{\nu}F_{\mu\nu}$. 
}
\begin{eqnarray}
 W_m(C)     &=& \exp(2\pi i \sum_{x,\mu}k_{x,\mu}N_{x,\mu}) ,
 \label{monopole dominance-1}     
\\
N_{x,\mu}  &=& \sum_{x'}\Delta_L^{-1}(x-x')\frac{1}{2}
\epsilon_{\mu\alpha\beta\gamma}\partial_{\alpha}
S^J_{s'+\hat{\mu},\beta\gamma}, 
\quad
\partial'_{\beta}S^J_{x,\beta\gamma} = J_{x,\gamma} ,
\label{monopole dominance-2}
\end{eqnarray}
where $N_{x,\mu}$ is defined through the external source $J_{x,\mu}$ which is used 
to calculate the static potential:
$\partial'$ denotes the backward lattice derivative
$\partial_{\mu}^{'}f_x=f_x-f_{x-\mu}$,  $S^J_{x,\beta\gamma}$ denotes a surface bounded by the 
closed loop $C$ on which the electric source $J_{x,\mu}$ has its support, and $\Delta_L^{-1}(x-x')$ 
is the inverse Lattice Laplacian. 
We obtain the string tension by evaluating the average of (\ref{monopole dominance-1}) from the 
generated configurations of the monopoles $\{k_{x,\mu}\}$.
Note that (\ref{monopole dominance-1}) is a gauge-invariant operator, since the monopole current defined
by (\ref{cfn-conti-20}) is a gauge-invariant variable.

\section{Numerical results}

First of all, we generate the configurations of SU(2) link variables 
$\{ U_{x,\mu} \}$, $ U_{x,\mu}= \exp [ - ig\epsilon {\bf A}_\mu(x) ]$,
using the standard Wilson action based on the heat bath method. 
Next, we generate the configurations of the color vector field $\{{\bf n}_x\}$ 
according to the (\ref{reduction_n}) together with the configurations of SU(2) 
link variables $\{U_{x,\mu}\}$. 
Then we can construct $\{V_{x,\mu}[U,{\bf n}]\}$ from (\ref{cfn-mono-4})
and $\{k_{x,\mu}\}$ from (\ref{cfn-conti-20}).

We calculate the respective potential $V_i(R)$ from the respective average $\left<W_i(C)\right>$:
\begin{eqnarray}
V_i(R) = -\log \left\{ \left< W_i(R,T) \right>/\left<W_i(R,T-1)\right> \right\} 
\quad (i=f, Abel, m)
\label{monopole dominance-3}
\end{eqnarray}
where $C=(R,T)$ denotes the Wilson loop $C$ with side lengths $R$ and $T$.  

\begin{figure}[h]
\vspace*{5mm}

\centerline{
\epsfxsize=0.7\textwidth
\epsfbox{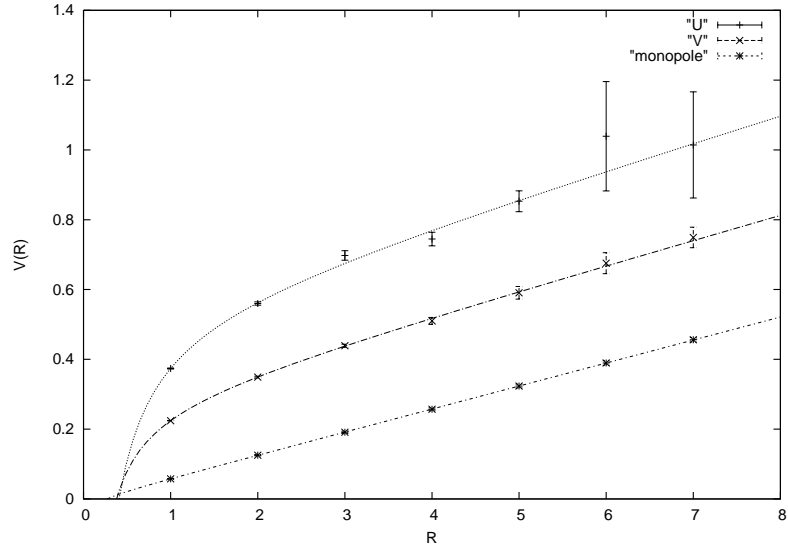}
}
\caption{The full, Abelian and monopole potentials as functions of $R$ at 
$\beta=2.4$ on $16^4$ lattice.}
\label{fig:potential}
\end{figure}

The numerical simulations are performed on an $16^4$ lattice at $\beta=2.4$ by 
thermalizing 3000 sweeps. In particular, we have used 100 configurations for the calculation of the 
full and Abelian potential and 50 configurations for the monopole potential in each 
case with 100 iterations. In order to obtain the full SU(2) and Abelian results, especially, we used the 
smearing method \cite{albanese87} as noise reduction techniques.
Fig.\ref{fig:potential} shows all potentials as functions of $R$.
The obtained numerical potential is fitted to a linear term,  Coulomb term and a constant term: 
\begin{eqnarray}
V_i(R) = \sigma_i R -  \alpha_i/R +c_i ,
\label{monopole dominance-4}
\end{eqnarray}
where $\sigma$ is the string tension, $\alpha$ is the Coulomb coefficient, 
and $c$ is the coefficient of the perimeter decay:
$\left<W_i(R,T)\right> \sim \exp [-\sigma_i RT -c_i(R+T)+\alpha_i T/R + \cdots]$. 
The results are shown in 
 Table~\ref{strint-tension-1}.

\begin{table}[h]
\caption{String tension and Coulomb coefficient}
\label{strint-tension-1}
\begin{center}
\begin{tabular}{lll}\hline
             & $\sigma$    & $\alpha$ \\ \hline
U (full)     & 0.075(9)    & 0.23(2)  \\
V (Abelian)  & 0.070(4)    & 0.11(1)  \\
monopole     & 0.066(2)    & 0.003(7) \\
\hline
\end{tabular}
\end{center}
\end{table}
\vspace{-.5cm}

We find that the Abelian part $\sigma_{Abel}$ reproduces 93$\%$ of the full string tension $\sigma_f$
and the monopole part $\sigma_m$ reproduces 94$\%$ of $\sigma_{Abel}$. 
Thus, we have confirmed the abelian and the monopole dominance in the string tension in our framework.
In general, the monopole part does not include  the Coulomb term and hence the potential 
is obtained to an accuracy better than the full potential.

\section{Conclusion}

In this talk, we have proposed a new formulation of the NLCV of Yang-Mills theory, which was 
once called the CFN decomposition.  
The Abelian and monopole dominances in the string tension has been shown anew 
in the gauge invariant way, whereas they have been so far shown only in a special gauge fixing 
called  MA gauge which breaks the color symmetry explicitly. 

\section{Acknowledgments}

The numerical simulations have been done on a supercomputer (NEC SX-8) at Research Center for 
Nuclear Physics (RCNP), Osaka University. 
This work is in part supported by the Large Scale Simulation Program No.08-16 (FY2008) and 
No.09-15 (FY2009) of High Energy Accelerator Research Organization (KEK).
This work is financially supported in part by Grant-in-Aid for 
Scientific Research (C) 21540256 from Japan Society for the Promotion of 
Science (JSPS).


\end{document}